\documentclass{icrc29}
\usepackage{graphicx,amssymb,amsmath,times}
\begin{document}
\title[Spectrum by Tunka ...]{Cosmic Ray Energy Spectrum and Mass Composition
from $10^{15}$ to $10^{17}$~eV by Data of the Tunka EAS Cherenkov Array} 
\author[N.~M.~Budnev et al.] {N.~M.~Budnev$^b$, 
 D.~V.~Chernov$^a$,
 O.~A.~Gress$^b$, 
 T.~I.~Gress$^b$,
 E.~E.~Korosteleva$^a$, 
        \newauthor
 L.~A.~Kuzmichev$^a$, 
	 B.~K.~Lubsandorzhiev$^c$, 
 L.~V.~Pankov$^b$,
 \framebox{Yu.~V.~Parfenov$^b$}, 
        \newauthor
 V.~V.~Prosin$^a$, 
 Yu.~V.~Semeney$^b$, 
 Ch.~Spiering$^d$, 
 R.~P.~Wischnewski$^d$,
 I.~V.~Yashin$^a$,
 \\
 (a) Skobeltsyn Institute of Nuclear Physics, Moscow State University, Russia     
 \\ 
 (b) Institute of Applied Physics, Irkutsk State University, Irkutsk, Russia  
 \\ 
 (c) Institute of Nuclear Research, Russian Academy of Science, Moscow, Russia  
 \\ 
 (d) DESY, Zeuthen, Germany      
        }
\presenter{Presenter: V.V. Prosin (prosin@dec1.sinp.msu.ru), \  
rus-prosin-VV-abs2-he12-oral}

\maketitle

\begin{abstract}

   We present results of an improved analysis of the experimental data of 
the EAS Cherenkov array Tunka-25. 
   A new function to fit the Cherenkov
   light lateral distribution LDF at core distances from 0 to 350 m has 
 been developed on the base of CORSIKA simulations and applied to the analysis
 of Tunka data.
   Two methods to estimate the EAS maximum position have been used. 
 The one is based on the pulse FWHM,  the other on the light LDF.  
   We present the primary energy spectrum in the energy range $10^{15}$ -
   $10^{17}$ eV.
   The use of the depth of the EAS maximum to determine the mean mass
   composition is  
 discussed.

\end{abstract}

\section*{Tunka-25: Experiment and Simulatons}
The Tunka EAS Cherenkov array is located in Tunka Valley, at an altitude of 
$675$~m a.s.l., and was described in Ref.~\cite{1}. 


Extensive air showers in an energy range $3\cdot 10^{14}$
to $2\cdot 10^{16}$ eV have been simulated with CORSIKA. 
The total amount of 600 CORSIKA events has been simulated for 
primary protons and iron nuclei and for three zenith angles 0$^\circ$,
15$^\circ$ 
and 25$^\circ$. The LDF consists of two branches: one almost exponential from
the core 
to some distance $R_{kn}$, the other following a power law from $R_{kn}$ up to
350 m:   
$$
 Q(R)=\left\{
\begin{array}{rl}
 Q_{kn}\cdot exp((R_{kn}-R)\cdot (1+3/(R+3))/R_0), & for R< R_{kn} \\
 Q_{kn}\cdot (R_{kn}/R)^{b}, &              for R\geq R_{kn}
\end{array} \right. \eqno (1)
$$
 $$R_0=10^{2.95-0.245P},  m$$
 $$R_{kn}=155-13P,  m$$
 $$ b=1.19+0.23P $$
The main difference to the expression suggested in \cite{2} is the use of
variable power law index for the second branch. All three partial parameters
defining the LDF 
shape, $R_0$, $R_{kn}$ and $b$, are strict functions of a single 
steepness parameter
$P$, defined as the ratio of Cherekov light fluxes at core distances 100 and
200 m:  $P = Q(100)/Q(200)$.

Examples of fitting the simulated LDF with 
equation 1 for different parameters $P$ are shown in fig.1.
The parameter $P$ and the light flux at a fixed core distance 
175~m, $Q_{175}$, have been
determined for every simulated event. 

The essential correlations of parameter pairs $E_0$ and $Q_{175}$, $E_0$ and
$H_{max}$, $H_{max}$ and $P$, and the standard deviations of their distibutions,
separately for $p$  and $Fe$, have been extracted from this CORSIKA simulation 
in order to
use them in a special code called "model of experiment". Two intermediate nuclear
groups, corresponding to primaries $He$ and $CNO$ have been added to the
"model" using an interpolation of parameters, obtained for $p$ and $Fe$. 
The real
geometry of the array and the response of every detector and it's fluctuations
are taken into account in the model. The
"model of experiment" permits a fast generation of
$10^4 - 10^5$ artificial events, distributed with a
realistic energy spectrum, for different 
assumptions concerning the primary mass composition. It allows
to analyse the array response,
including the analysis of details of the program of EAS parameter
reconstruction, 
the efficiency of registration of EAS, the threshold of the total array, errors
and 
possible  distortions of the reconstructed energy $E_0$, and the
depth of the shower maximum $X_{max}$.      
 
The "model of experiment", assuming a complex mass composition
(\mbox{p:He:CNO:Fe=0.25:0.25:0.25:0.25}), 
has shown that 
for energies $ > 3\cdot 10^{15}$ eV,
our procedures of relative calibration 
of the detectors and the EAS parameter reconstruction result in 
accuracies of primary energy determination of $\sim 15$\%,
of the core position of $\sim$5 m and 
of $X_{max}$ of $\sim$30 g/cm$^2$.
For energies below  $3\cdot 10^{15}$ the errors are larger. 
The threshold of data acquisition with 100\% (50\%) efficiency 
is $8\cdot 10^{14}$ eV ($5\cdot 10^{14}$ eV).  
Some other conclusions of the "model" will be discussed later.  

\section*{Primary Energy Spectrum}

\begin{figure}[t]
\includegraphics*[width=0.5\textwidth,angle=0,clip]{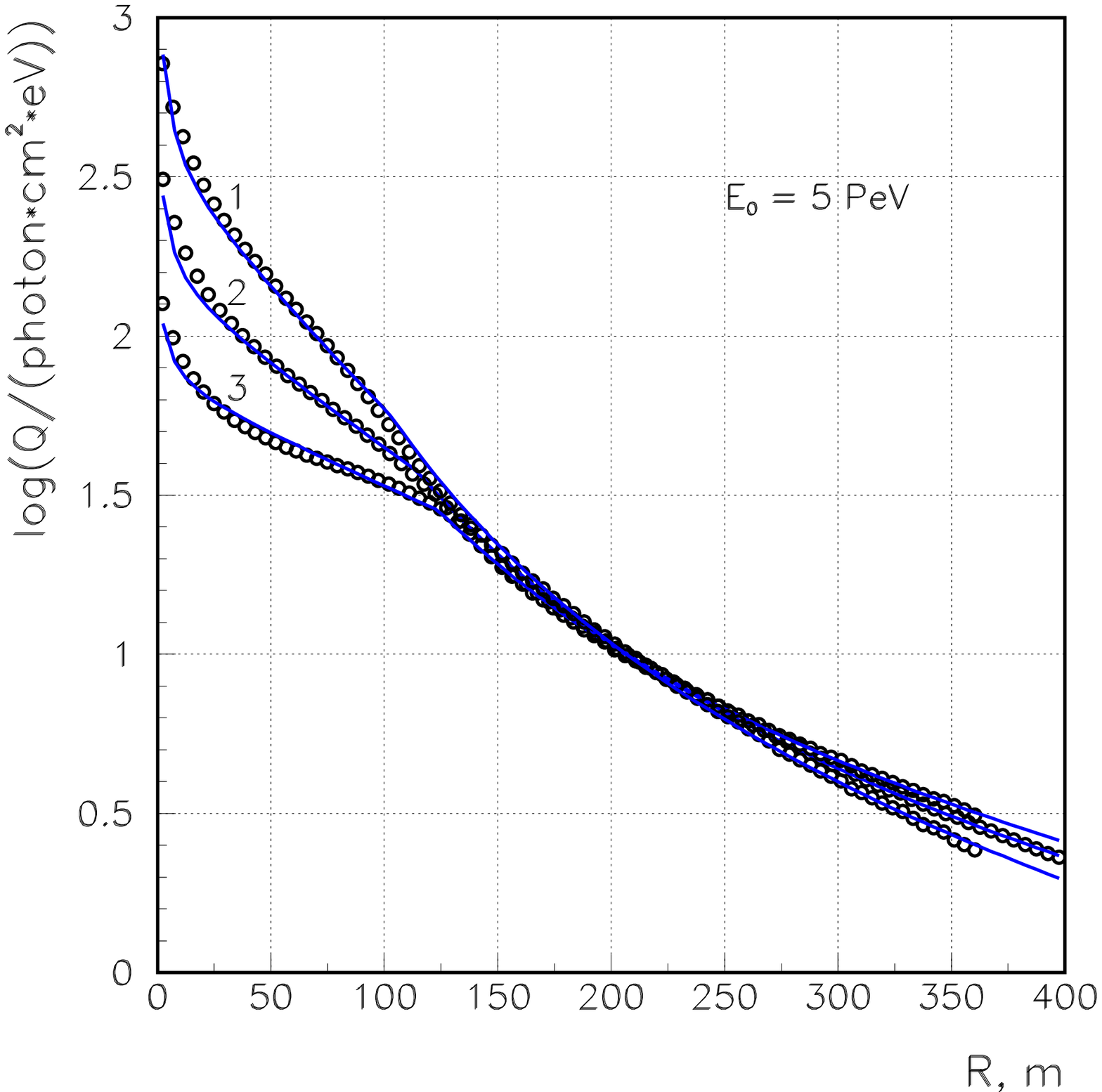}
\hfill
\includegraphics*[width=0.5\textwidth,angle=0,clip]{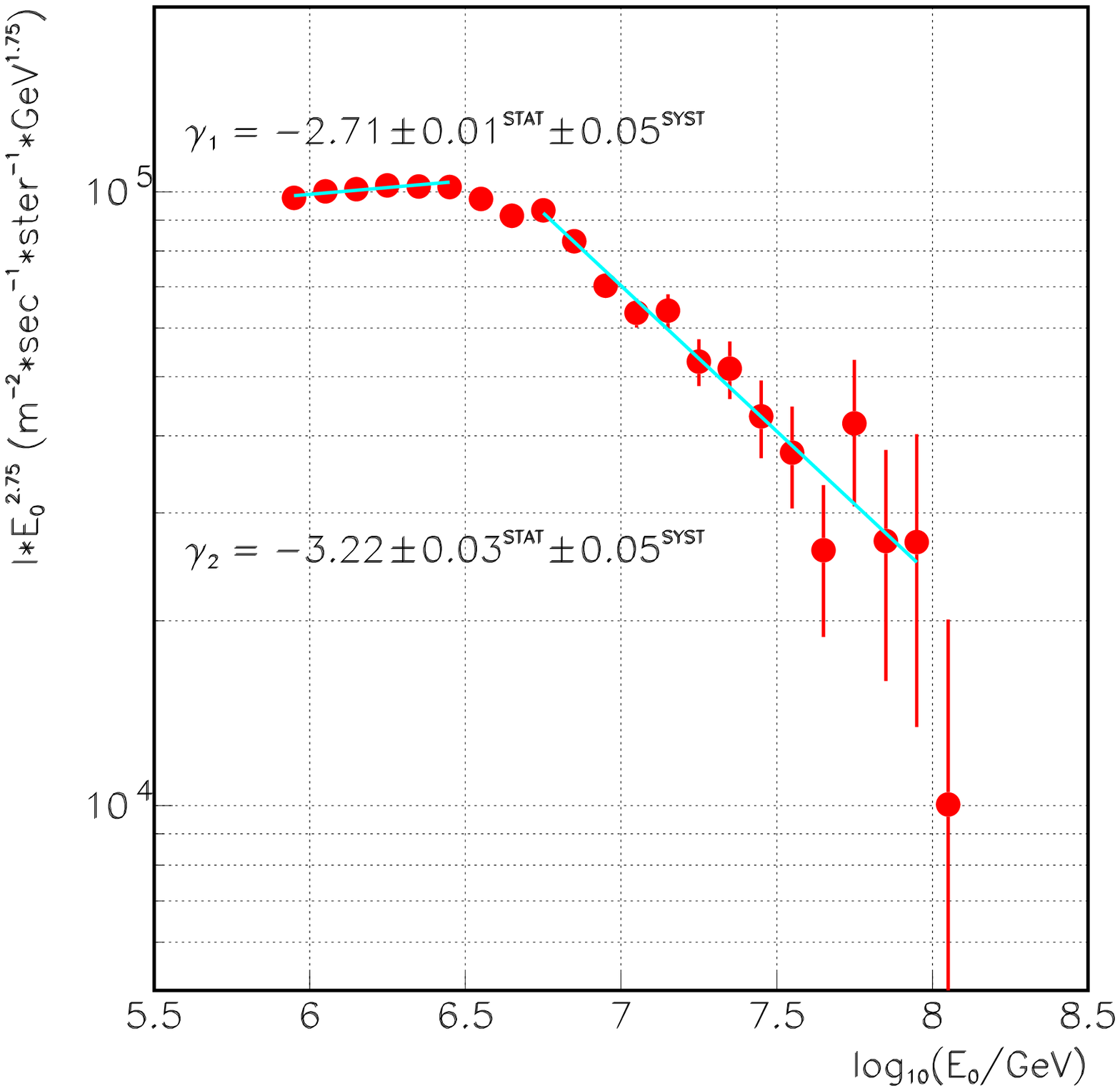}
\vspace*{8pt}
\parbox[t]{0.48\textwidth}{\caption{CORSIKA: 
EAS Cherenkov light LDFs and fitting functions. 
1 - $P$ = 5, 2 - $P$ = 4, 3 - $P$ = 3}}
\hfill
\parbox[t]{0.48\textwidth}{\caption{EXPERIMENT: 
Differencial energy spectrum.}}
\end{figure}


The primary energy 
$E_0$~[TeV] has been obtained from 
$Q_{175}$~[$\mathrm{photon}\cdot\mathrm{cm}^{-2}\cdot\mathrm{eV}^{-1}$] with 
the relation: $$E_0=400\cdot{Q}_{175}^{0.95}.$$ The absolute energy 
calibration is based on the results obtained with the QUEST experiment \cite{3}.

Figure 2 presents the differential energy spectrum, derived from data taken in
$300$~hours, spread over about $50$ clear,  
moonless nights, with a trigger rate above $1.8$~Hz . 
To construct a spectrum,
showers with zenith angles $\theta\leq25^{\circ}$ and a core position inside the
geometrical area of the array have been selected. 

We note that particularly at the lowest and the highest energies,
the new fit functions performs significantly better than that
used in \cite{4}.

\section*{Depth of the EAS Maximum}


The lateral and the time distributions of the
Cherenkov light provide two independent 
methods to estimate the depth of the EAS maximum. 
The first is the measurement of the LDF
steepness $P$, which is related to the distance to the 
shower maximum by the expression 
$H_{max}$ (in~[km]): $H_{max}=17.63-0.0786*(P+8.916)^2$. 
This relation is almost independent of other details of the
simulation: energy,
sort of nucleus, zenith angle and model of hadron interaction. 
The distribution of
$H_{max}$ for a fixed $P$ has a standard deviation of only 0.3\,km.

The depth of the EAS maximum $X_{max}$ is derived from $H_{max}$ using the
atmosphere parametrs consistent with really observed (the mean temperature
during the observations is -20$^\circ$C). 
The "model of experiment" gives an error of 30~g/cm$^2$ 
for the experimental depth of the EAS maximum $X_{max}$.

The second method is measuring the Cherenkov pulse full width on half maximum
(FWHM). In accordance with CORSIKA simulations as described above, 
the FWHM~[ns] at distances larger than $200$~m from the   
EAS axis is related to the relative position of the EAS maximum: 
$\Delta{X}=X_0/\cos\theta-X_{\max}$~[g/cm$^2$], -- where $X_0$ is the total
depth  
of the atmosphere and $\theta$ is the zenith angle of the shower. The relation
between FWHM and $\Delta{X}$
depends only on the distance to the EAS axis, and is almost independent on the
other details of the simulation: energy, sort of nucleus and model of hadron
interaction. For example, for a distance of 
$250$~m, one obtains 
$\Delta{X}=1677+1006\cdot\log_{10}(\mathrm{FWHM})$. 
The underlying theoretical uncertainties for this method
are smaller than for the first one. 

Figure~3 presents the mean depth of the EAS maximum, derived with the two 
methods described above, as a function of primary energy. 
The "model of experiment"
shows that there is no influence of the array threshold on the obtained
$X_{max}$, starting from an energy $1.5\cdot 10^{15}$ eV. It is seen from fig.3 that
the threshold of the FWHM method is higher than that of the LDF steepness method, 
but the mean depths, obtained with the two different methods are in good
agreement. Both methods result in large fluctuations of mean points at low
statistics (high energy), as expected for a very asymmetric distribution
like that of  $X_{\max}$. 

Experimental  $X_{\max}$ distributions for 
EAS in the energy range 3 -- 10 PeV are shown in fig 4. 
All the values of $X_{\max}$ are re-normalized to a fixed energy of
$5\cdot 10^{15}$~eV, using an
experimental elongation rate of ($97\pm 3$) g/cm$^2$.
The experimental distribution is compared to distributions simulated 
with the "model
of experiment", for different assumptions on primary composition, and
re-normalized in the same way as the experimental one. 
The agreement with experimental data is better
for the complex composition (\mbox{p:He:CNO:Fe=0.25:0.25:0.25:0.25}), 
than for pure $p$ or $Fe$. 
So the intermediate mass composition is more probable, but it
should be noticed that the difference
between simulated and experimental standard deviation is significant
only for pure $p$ or $Fe$. For a wide range of intermediate
compositions, mixtures of mass components lead to standard
deviations similar 
to those marked as "complex" in the figure.

The mean value of the simulated distribution depends on 
hadron interaction model. In fig.~3,   
$X_{\max}$ vs. energy is displayed
for two models QGSJET-01 \cite{5} and QGSJET-II \cite{6}. 
All other models used in CORSIKA yield lines between these borders
\cite{7}. 

One sees, that a conclusion about the mean mass composition strongly depends
on the chosen hadron interaction model. To solve the problem with the model the
measurement of energy dependence of $X_{\max}$ in much wider energy
range is very essential. This is one of the reasons of our decision to spread
our measurement to lower energy with the new version of Tunka-25 Cherenkov light
detectors supplied with spheric mirrors of 1 m diameter. To the other hand the
new experiment Tunka-133 \cite{8} with effective area 10 times higher than that
of Tunka-25 is under construction in Tunka valley. 

\begin{figure}[t]
\includegraphics*[width=0.5\textwidth,angle=0,clip]{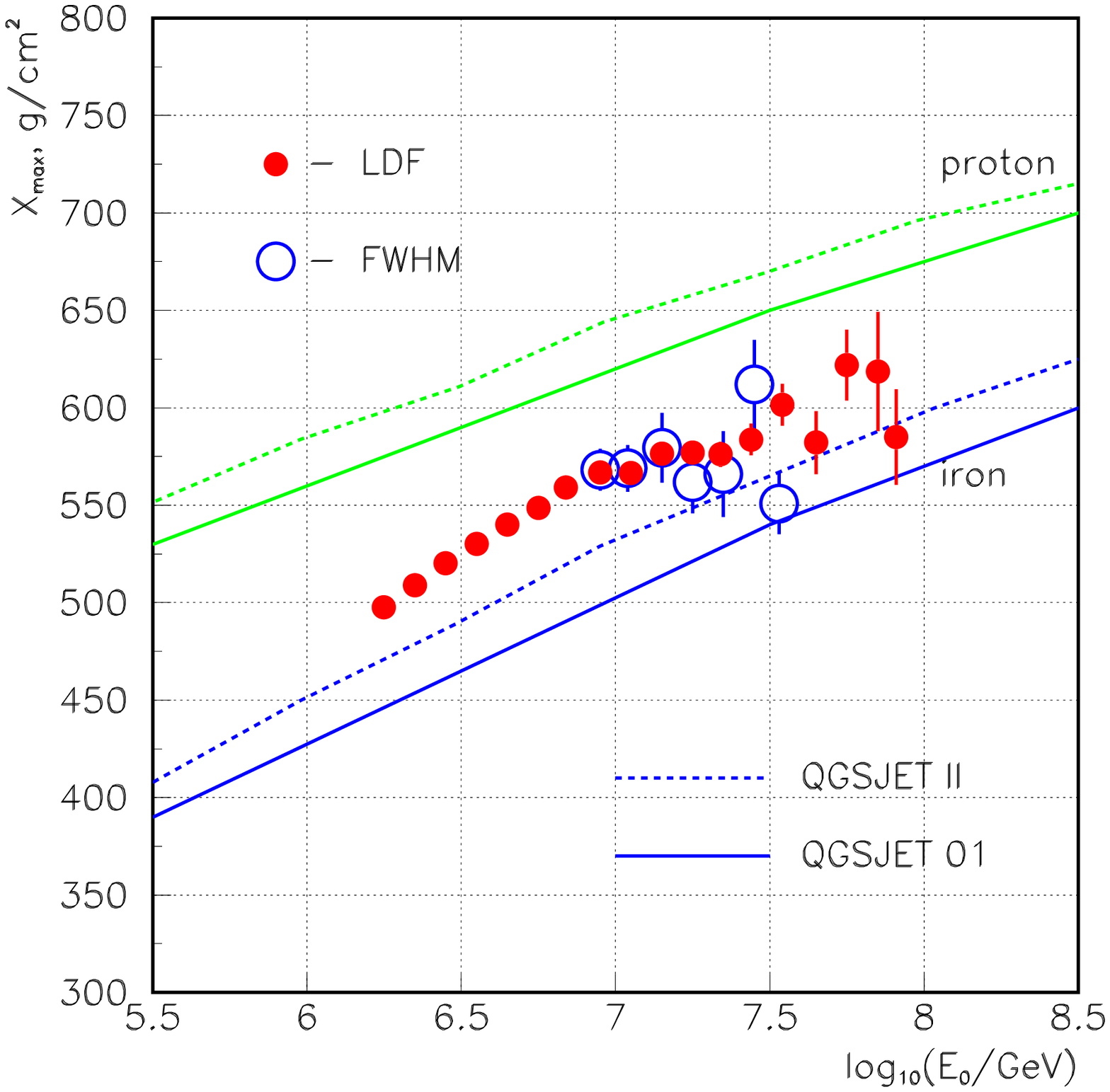}
\hfill
\includegraphics*[width=0.5\textwidth,angle=0,clip]{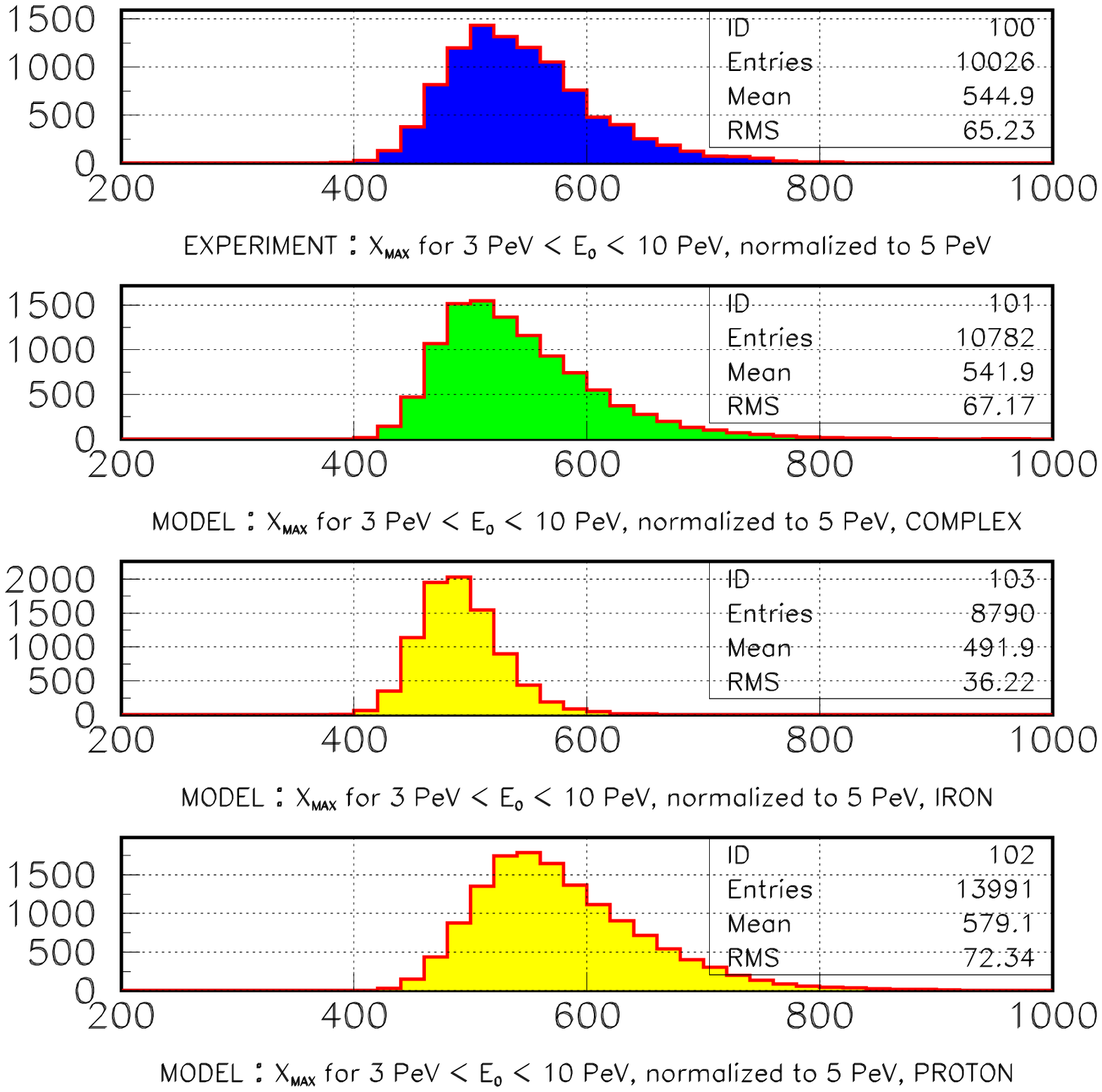}
\vspace*{8pt}
\parbox[t]{0.48\textwidth}{\caption{ 
Mean depth of EAS maximum vs energy.}}
\hfill
\parbox[t]{0.48\textwidth}{\caption{$X_{max}$ distribution.
}}
\end{figure}

\section*{Acknowledgements}
      
The authors are thankful to professor Gianni Navarra and the EAS-TOP Collaboration for 
the opportunity to carry out the calibration experiment QUEST at the EAS-TOP 
array. This work is supported by the Russian Fund of Basic Research 
(grants: 03-02-16660, 05-02-04010 and 05-02-16136).


\begin{thebibliography}{99}

\bibitem{1}
N.~M.~Budnev et al., 27th ICRC, Hamburg (2001) 2, 581 .
\bibitem{2}
E.~E.~Korosteleva, L.~A.~Kuzmichev, V.~V.~Prosin and the EAS-TOP Collaboration,
28th ICRC, Tsukuba (2003) 1, 89.
\bibitem{3}
E.~E.~Korosteleva, L.~A.~Kuzmichev, V.~V.~Prosin and the EAS-TOP Collaboration,
IJMPA (2005) in print, astro-ph/0411216.
\bibitem{4}
D.~V.~Chernov et al., IJMPA (2005) in print, astro-ph/0411139.
\bibitem{5}
N.~N.~Kalmykov et al., Nucl. Phys. B (Proc. Suppl.), 52B (1997) 17.
\bibitem{6}
S.~Ostapchenko, Nucl. Phys. B (Proc. Suppl.), (2005) in print, hep-ph/0501093.
\bibitem{7}
D.~Heck, Talk presented at the VIHKOS CORSIKA School 2005, Lauterbad. 
\bibitem{8}
D.~V.~Chernov et al., IJMPA (2005) in print, astrop-ph/0411218.

\end{thebibliography}
\end{document}